\documentclass[twocolumn,showpacs,superscriptaddress,prl]{revtex4-1}

\usepackage[dvips]{graphicx}	
\usepackage{dcolumn, bm, color, amsmath}
\usepackage{amssymb}

\begin{document}

\title{Unconventional gap structures and the intermediate mixed state: a vortex lattice study of the noncentrosymmetric superconductor BiPd}

\author{A.~S.~Cameron}
\affiliation{School of Physics and Astronomy, University of Birmingham, Edgbaston, Birmingham, B15 2TT, UK.}
\affiliation{Institut f\"ur Festk\"orper- und Materialphysik, Technische Universit\"at Dresden, D-01069 Dresden, Germany.}
\author{L. Lemberger}
\affiliation{School of Physics and Astronomy, University of Birmingham, Edgbaston, Birmingham, B15 2TT, UK.}
\affiliation{Institut Laue Langevin, 71 Avenue des Martyrs, F-38000 Grenoble cedex 9, France.}
\author{R. Riyat}
\affiliation{School of Physics and Astronomy, University of Birmingham, Edgbaston, Birmingham, B15 2TT, UK.}
\author{A. T. Holmes}
\affiliation{School of Physics and Astronomy, University of Birmingham, Edgbaston, Birmingham, B15 2TT, UK.}
\affiliation{European Spallation Source ERIC, P.O. Box 176, SE-221 00, Lund, Sweden.}
\author{Y. S. Yerin}
\affiliation{CNR-SPIN, via del Fosso del Cavaliere, 100, 00133 Roma, Italy.}
\author{A. D. Hillier}
\affiliation{ISIS Pulsed Neutron and Muon Source, STFC Rutherford Appleton Laboratory, Harwell Campus, Didcot OX11 0QX, United Kingdom.}
\author{B. Joshi}
\affiliation{Tata Institute of Fundamental Research, Mumbai-40005, India.}
\author{S. Ramakrishnan}
\affiliation{Tata Institute of Fundamental Research, Mumbai-40005, India.}
\author{J. Gavilano}
\affiliation{Laboratory for Neutron Scattering and Imaging (LNS), Paul Scherrer Institute (PSI), CH-5232 Villigen, Switzerland.}
\author{C.~D.~Dewhurst}
\affiliation{Institut Laue Langevin, 71 Avenue des Martyrs, F-38000 Grenoble cedex 9, France.}
\author{E.~M.~Forgan}
\affiliation{School of Physics and Astronomy, University of Birmingham, Edgbaston, Birmingham, B15 2TT, UK.}
\author{E.~Blackburn}
\email{elizabeth.blackburn@sljus.lu.se}
\affiliation{School of Physics and Astronomy, University of Birmingham, Edgbaston, Birmingham, B15 2TT, UK.}
\affiliation{Division of Synchrotron Radiation Research, Lund University, SE-22100 Lund, Sweden.}

\begin{abstract}
We report on neutron scattering measurements on the vortex lattice of the noncentrosymmetric superconductor BiPd. We observe the existence of the intermediate mixed state, a region where Meissner and vortex lattice phases coexist, which is a feature of low $\kappa$ Type-II superconductors. Following this, we obtain an estimate of the value of $\kappa$ using the extended London model, which confirms the expectation that $\kappa$ should be small. Finally, we find that the temperature dependence of the vortex lattice form factor fits well to a model designed to describe singlet-triplet mixing in non-centrosymmetric superconductors, which may shed light on the question of the gap structure in BiPd.
\end{abstract}



\date{\today}

\maketitle

\section*{Introduction}

Superconductivity in non-centrosymmetric (NCS) materials is of particular interest, as it has been predicted that the breaking of spatial inversion symmetry by the crystal lattice could lead to unconventional pairing symmetries where the mixing of singlet and triplet order parameters is possible~\cite{Sig91, Sig07, Gor01, Bau04, Fri04,  Bau12}. Following the discovery of superconductivity in the NCS system CePt$_3$Si~\cite{Bau04}, investigations into these materials have flourished, and many more NCS superconductors have now been reported~\cite{Smi17}. Broken inversion symmetry leads to anisotropic spin-orbit coupling (ASOC), which removes the spin degeneracy of the electronic bands~\cite{Rashba60}, leading to spin-split Fermi surfaces. Combined with their unique pairing state, which may itself generate anisotropic gaps and accidental nodes~\cite{Hay06}, this results in a variety of unusual phenomena~\cite{Karki10, Chen11, Takimoto09, Samokhin04, Yua06}, including spontaneous magnetisation at twin boundaries~\cite{Arahata13} and helical phases of the order parameter~\cite{Mineev94, Kau05}. From the perspective of the vortex lattice (VL), there is a predicted stabilisation of the Fulde-Ferrell-Larkin-Ovchinnikov state~\cite{Tanaka07}, a weakened paramagnetic limiting response and the emergence of the anomalous magnetoelectric effect~\cite{Edelstein95, Mineev05, Fujimoto05, Mineev11, Hiasa08}. Theoretical studies of the structure of vortices and the VL have predicted an altered structure~\cite{Yip05, Hay06} and the emergence of tangential components of the magnetic field~\cite{Mas06, Lu09, Kas13}. Furthermore, experimental evidence for unusual behaviour of the VL in the NCS material Ru$_7$B$_3$ has previously been reported~\cite{Cam19, Cam23}, as well as evidence for singlet-triplet mixing in the order parameter of this and several other NCS superconductors~\cite{Cam22, Yua06}, indicating this is a rich field of study with many interesting phenomena predicted.
\begin{figure*}
	\includegraphics[width=1\linewidth]{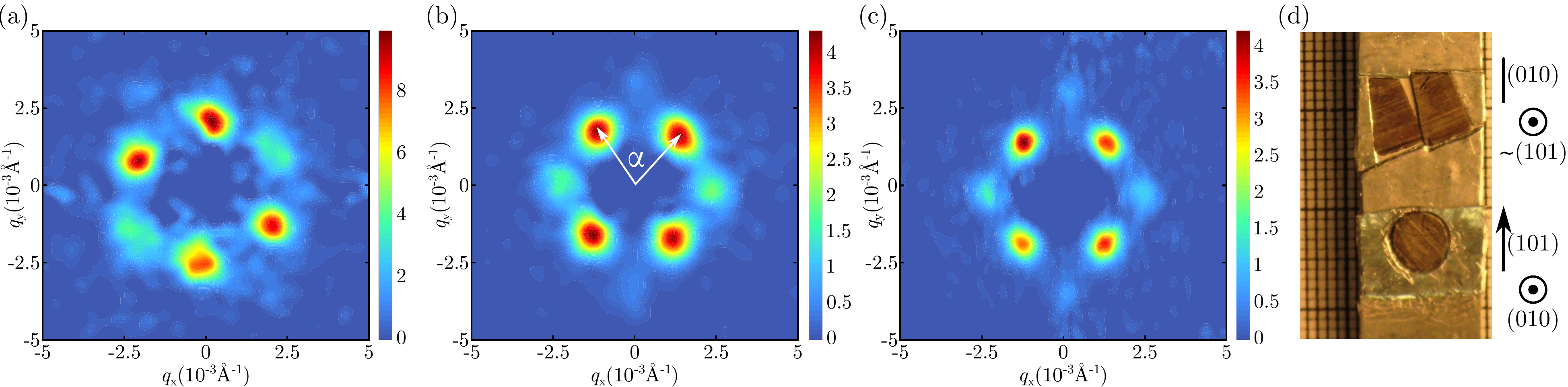}
	\caption{(Color online) (a-c) Diffraction patterns at around 200~G for fields applied along different crystallographic directions. (a) 220~G with $B \parallel (010)$, at 0.05~K. (b) 200~G with $B \parallel \sim$(101), at 0.1~K. (c) 210~G with $B \parallel (101)$, at 0.1~K. (d) A photograph of the samples with the $\sim$(101) (top) and (010) (bottom) orientations perpendicular to the sample holder. The angle $\alpha$ between two scattering vectors of the vortex lattice is given in panel (b).}
		\label{Graph1}
\end{figure*}

$\alpha$-BiPd, which we refer to in this paper simply as BiPd, is an NCS material, forming a monoclinic structure below 210$^{\circ}$C, with space group P2$_1$~\cite{Bha79}. It undergoes a superconducting transition at 3.8 K, but there is some contention over the upper critical field, as different measurements have produced wildly different results. Electrical resistivity measurements by Joshi \emph{et al.}~\cite{Jos11} found a lower critical field of 123~G and an upper critical field of 7000~G for fields applied along the (010) direction, with a moderate anisotropy of not greater than $30 \%$ for fields applied along the other crystallographic axes.  From this, the GL parameter was estimated to be $\kappa \sim 7.6$~\cite{Jos11}, which would make this a moderate Type-II superconductor. However, magnetisation and heat capacity measurements~\cite{Pee14, Sun15} gave values of the upper critical field an order of magnitude smaller than this. To date, there is not a conclusive consensus on the origin of this discrepancy, although the presence of filamentary superconductivity has been proposed~\cite{Pee14, Ram17}.

Fermi surface measurements point to the existence of multiple bands \cite{Ram15}, which is not surprising as at the very least NCS materials have spin-split bands due to the ASOC. The gap structure of BiPd is the subject of a complex debate. Some heat capacity measurements by Ramakrishnan \textit{et al.}~\cite{Ram17} were best fit to a two-gap model; an anisotropic single-gap model did not appear to fit the data.  Point contact spectroscopy measurements \cite{Min12} also supported a two-gap model. The presence of a conductance peak at zero bias in these data was attributed to Andreev bound states~\cite{Min12, Kas00}, which indicated an unconventional nature for the order parameter. However, bulk measurements of magnetisation, resistivity and specific heat reported by Peets \cite{Pee14} can all be explained using conventional \textit{s}-wave behaviour, although Peets notes that the upward curvature of the upper critical field is suggestive of possible multi-band superconductivity.   An STM study by Sun \emph{et al.}~\cite{Sun15} saw no evidence for anisotropy in the gap~\cite{Sun15}, but other STM measurements show signs of unconventional superconductivity~\cite{Ram17}. Tunnel diode oscillator measurements of the penetration depth~\cite{Jia14} along different crystal axes were best fit to a two-band BCS model where one of the bands is anisotropic, although the authors concluded that such an order parameter could be due to the singlet-triplet mixing predicted for NCS superconductors. In short, while there is reasonable support for two-gap scenario, the question of anisotropy in the gap is contested.  Clearly, the situation is complex and requires further study.  In this paper, we provide additional evidence from values of the penetration depth, obtained via small angle neutron diffraction measurements of the vortex lattice.

\section*{Experimental Details}

Samples of BiPd with a superconducting transition temperature of $T_{\rm c} = 3.8$~K were grown by the process detailed in Ref.~\cite{Jos11}. They were cut into plates which were several mm along each side, but less than a mm thick, with chosen crystal directions perpendicular to the plane of the plates. These were then glued with Cytop\textregistered ~onto flat copper sample holders, with cadmium used to shield from the neutron beam those parts of the holder not containing the sample. Copper sample holders were chosen, instead of the aluminium typically used in neutron scattering, to avoid the aluminium becoming superconducting below 100 G \cite{Coc58}. Several samples were cut such that the (010) and (101) axes were accessible either with the sample plate perpendicular to the applied magnetic field and the neutron beam (Fig.\ref{Graph1}(d)), or where only a small rotation of the plate was needed.  For one sample, the cut was $13^{\circ}$ away from the (101) axis; we refer to this as $\sim$(101) in the rest of the article.

To measure the vortex lattice, small angle neutron scattering (SANS) measurements were carried out.  A vortex lattice (VL) was created by applying a magnetic field to the BiPd sample in the normal state and then cooling into the superconducting mixed state.  If the field (and the vortices) are nearly parallel to the neutron beam, the neutrons can diffract off the vortices, and we measure the first order Bragg peaks arising from the VL.

The SANS measurements were carried out on the SANS-I instrument \cite{Koh00} at the SINQ facility at the Paul Scherrer Institute (PSI) in Villigen, Switzerland, and the D33 instrument \cite{Dew08} at the Institut Laue-Langevin (ILL) in Grenoble, France. Incoming neutrons were velocity selected to have a wavelength between 10 and 22~\AA, {depending on the lattice spacing of the vortex lattice which changes as a function of field,} with a $\Delta \lambda / \lambda$ of around $10\%$. Diffracted neutrons were detected by a two-dimensional position-sensitive detector, and incoming neutrons were collimated such that the collimation length matched the sample-detector distance. Low temperatures were achieved using a dilution refrigerator insert in top-loading cryomagnets, which provide the horizontal field applied nearly parallel to the neutron beam. 

Measurements were taken by holding the applied field fixed with respect to the sample, and then rocking the sample and field together throughout the angles that fulfilled the Bragg condition for the first-order VL diffraction spots; examples of the resulting diffraction patterns are shown in Fig.~\ref{Graph1}. Temperature scans were taken on warming, after cooling the sample from above $T_{\rm c}$ in the required field. For these temperature scans, full rocking scans were not taken at each point, but instead extended counts at a rocking angle giving peak intensity at chosen Bragg reflections were carried out.  This was done to save time and maximise the signal-to-noise ratio.  To check that the angles chosen remained valid over the changes in temperature, at a few select temperatures full rocking scans were done.  This also allowed us to confirm that the widths of the rocking curves did not change as a function of temperature.  The samples were cooled in a constant field, rather than oscillating the field value during cooling, after initial tests found a superior VL quality with simple field cooling. The cooling rate was $\sim$ 0.1~K per minute. Background measurements were taken above $T_{\rm c}$, using the same set of rocking angles as the foreground measurements, and were then subtracted from the foregrounds measured inside the superconducting phase to leave only the signal due to the vortex lattice. Data reduction was performed using the GRASP~\cite{GRASP} software from the ILL, and diffraction patterns were treated with a Bayesian method for maximising the signal-to-noise of the small-angle diffraction data~\cite{Hol14}.

\section*{Results and Discussion}

Figure~\ref{Graph1} shows a representative selection of VL diffraction patterns obtained from fields applied along the (010), $\sim$(101), and (101) directions. These three diffraction patterns were taken at similar fields of 220~G, 200~G, and 210~G respectively, at a base temperature of between 0.05~K and 0.1~K. Each shows a slightly distorted hexagonal lattice, with the ratios of largest to smallest $q$ being 1.11, 1.09, and 1.12, respectively. Panel (d) is a picture of two of the samples used in these measurements, oriented such that the top sample in the picture has the magnetic field applied along the $\sim$(101) direction and the bottom sample has the field applied along the (010) direction. While there are measurable differences in the vortex structure dependent on the direction of applied field, as indicated by the different degrees of anisotropy, we did not observe a statistically significant variation in the vortex structure as a function of field or temperature.

\subsection{Intermediate mixed state}

Firstly, we report the observation of the intermediate mixed state (IMS) in part of the superconducting phase of BiPd. Within Ginzburg-Landau (GL) theory, the transition between Type-I and Type-II superconductors takes place when $\kappa = \lambda / \xi = 1 / \sqrt{2}$. If $\kappa$ is below this threshold, no mixed state exists, and only the Meissner state is observed in applied magnetic field (Type I).  For high-$\kappa$ superconductors (Type II), there is a clearly defined vortex lattice in the superconducting state between the Meissner state and the upper critical field. However, superconductors with $\kappa \gtrsim 1/\surd2$ have an additional phase between the Meissner and vortex-lattice phases. This is the intermediate mixed state, in which the superconducting state separates spatially into domains of vortex lattice state and Meissner state. This was first observed by Bitter decoration \cite{Tra67}, and has been well studied in niobium using neutron scattering ~\cite{Chr77,Bac19}, and for example more recently in ZrB$_{12}$ by Biswas \textit{et al.}~\cite{Bis20}.

The unusual behavior of vortex matter and the correspondingly more complex phase diagram of low-$\kappa$ superconductors has been predicted by various theoretical approaches as well. In particular, for the $s$-wave superconductor near the point $\kappa=1/\sqrt 2$, the $B$-$T$ phase diagram was studied in the framework of the GL theory, various tricritical points on it were determined, and regions with non-monotonic character of the inter-vortex interaction were characterized \cite{Luk'yanchuk}. A similar broadening of the region in the phase diagram between the Meissner and the mixed phases was also found in the framework of numerical solutions of microscopic Eilenberger equations for the isotropic $s$-wave gap \cite{Miranovic}.

\begin{figure}
\begin{center}
	\includegraphics[width=0.7\linewidth]{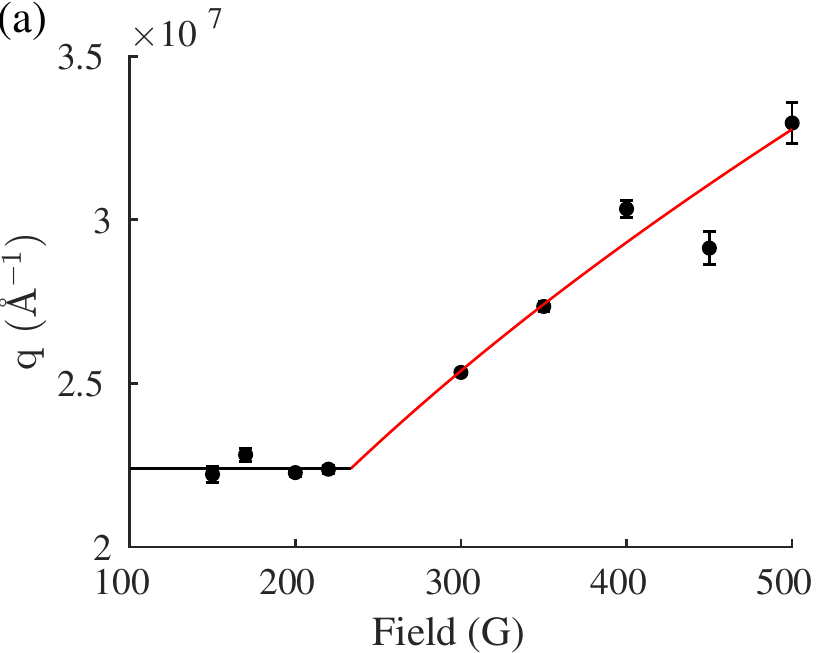}
	\includegraphics[width=0.7\linewidth]{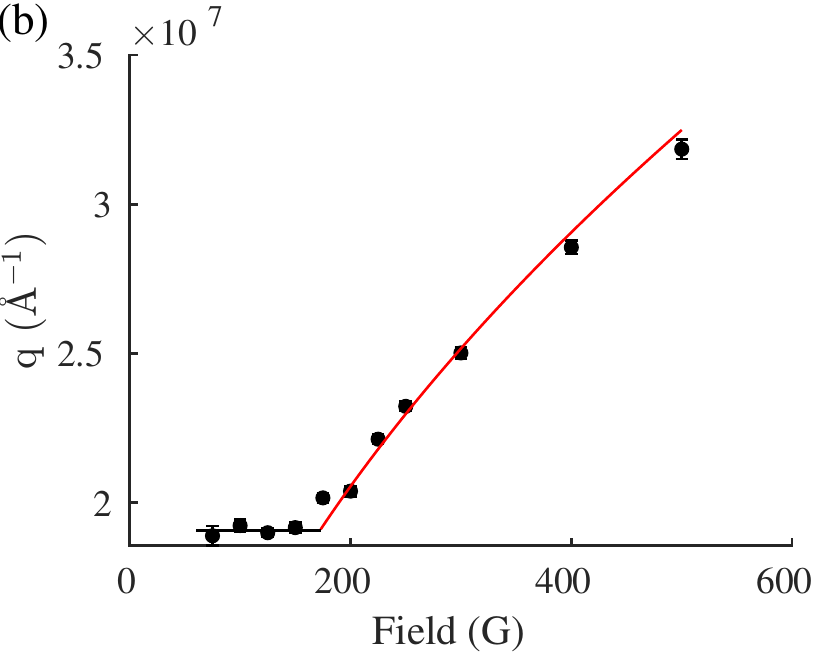}
	\includegraphics[width=0.7\linewidth]{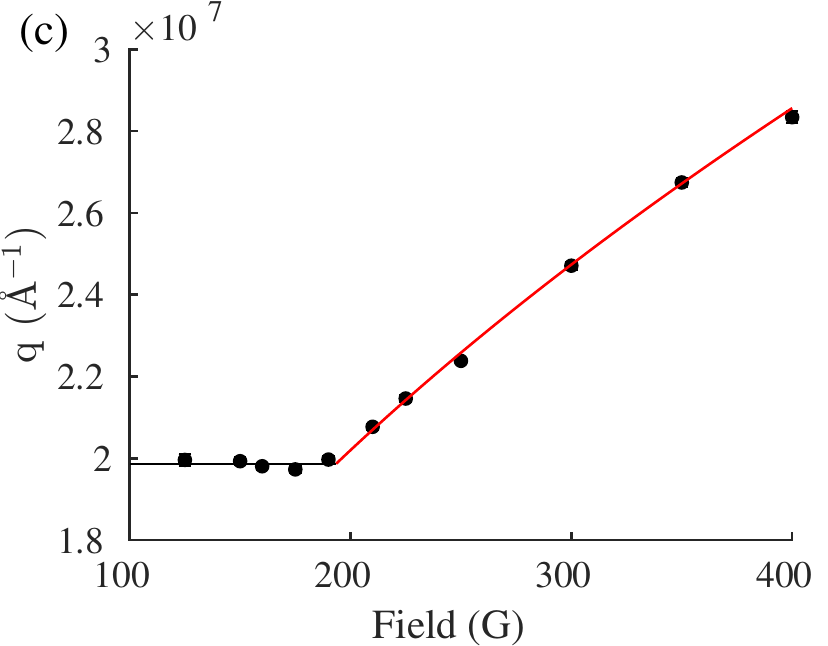}
\end{center}
	\caption{(Colour online) Length of the scattering vector, $q$, as a function of applied field along the (a) (010), (b) $\sim$(101), and (c) (101) directions respectively. { The flat lines (black) indicate the fixed vortex spacing characteristic of the intermediate mixed state.}  The field-dependent lines (red) are fits to the relationship described in the text.}
		\label{Graph2}
\end{figure}

In the mixed state, each vortex carries a quantum of flux, and so the density of vortices must be such as to match the magnetic flux density of the average induction in the sample. As the vortices are subject to a repulsive inter-vortex interaction, the unit cell size of the vortex lattice is set by the applied field. However, in the IMS, this relationship between the applied field and the separation of the vortices is broken.  The plane spacings become fixed as a function of applied field, being instead determined by the distance at which the inter-vortex interaction changes from repulsive to attractive~\cite{Bra11}. Hence in the IMS the VL regions have a flux density $B_0$ determined by their fixed spacing. These regions occupy a fraction of the sample equal to $B_0 \ / <B>$. The characteristic signature of the intermediate mixed state in neutron scattering is therefore a low-field region in which the scattering vector is independent of magnetic field. Figure~\ref{Graph2} is a plot of the magnitude of the scattering vector, $q$, for the stronger Bragg reflections in Fig.~\ref{Graph1}, in all three orientations of the magnetic field measured here. There is clearly a flat region at low field where the scattering vector is independent of magnetic field, and the high-field data follow the typical field-dependent relationship for a VL,
\begin{equation}
q = 2 \pi \sqrt{\frac{1}{\mathrm{sin(\alpha)}} \frac{B}{\Phi_0}},
    \label{Eqn_q}
\end{equation}
which arises geometrically from the constraint of one flux quantum per unit cell, where in this case $q$ is the length of the scattering vector of the four stronger spots in Fig.~\ref{Graph1}, $\alpha$ is the angle between two of them, illustrated in Fig.~\ref{Graph1}(b), and $\Phi_0$ is the flux quantum. Leaving $\alpha$ as a free parameter to fit the high-field data of Fig.~\ref{Graph2} returns opening angles between $70^{\circ}$ and $75^{\circ}$, which is consistent with the diffraction patterns of Fig.~\ref{Graph1}. If we define the critical field of the intermediate mixed state, $B_{\mathrm{IMS}}$, as the point where the flat region of $q$ (determined from the average of the low-field data points) meets the field-dependent region, we have $B_{\mathrm{IMS}} = 230$~G for $H \parallel (010)$, $B_{\mathrm{IMS}} = 170$~G for $H \parallel \sim$(101), and $B_{\mathrm{IMS}} = 190$~G for $H \parallel (101)$.

\begin{figure}
	\includegraphics[width=1\linewidth]{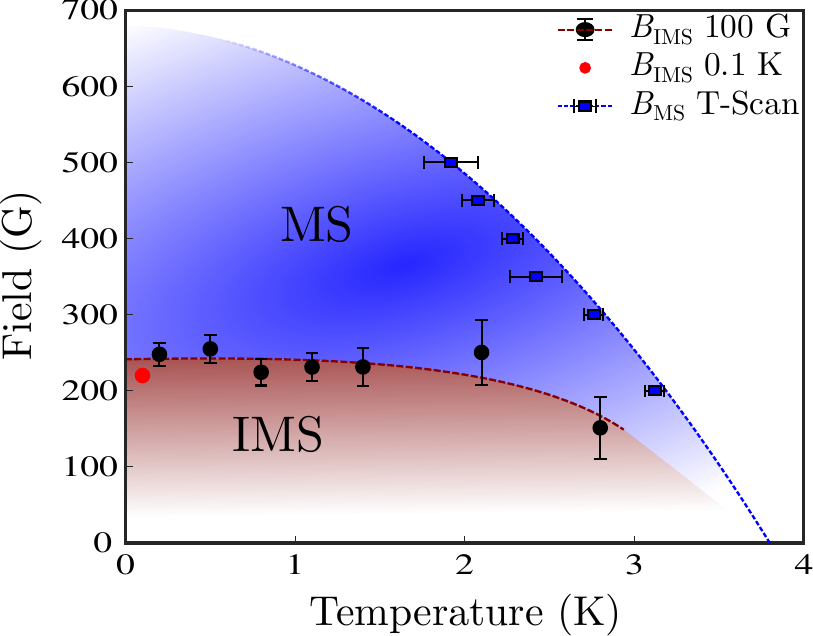}
	\caption{Vortex phase diagram of the vortex lattice phases in BiPd for magnetic field oriented along the $(010)$ direction. The upper critical field is indicated by squares, with the blue dashed line being a guide for the eye. The transition between the mixed state (MS) and intermediate mixed state (IMS) is indicated by black and red circles and the red dashed line, is labelled $B_{\rm IMS}$, and was determined by data from this study in a manner described in the text.}
		\label{Graph6}
\end{figure}

Figure~\ref{Graph6} is a phase diagram of the vortex lattice states in BiPd for field applied along the $(010)$ direction, which illustrates where the IMS sits in relation to the rest of the superconducting state, and is constructed from data in this study. Upper critical field points are determined from fits to the temperature dependence of the form factor, which will be discussed in a later section, and are shown by square markers. From this we estimate an upper critical field line, which is a guide to the eye. The transition points between the mixed state and intermediate mixed state were determined from SANS data by measuring the scattering vector of data taken within the IMS, at 100~G, and then calculating the corresponding mixed-state field in equation~\ref{Eqn_q}, and are shown in black circles. The data point at 0.1~K is also shown with a red circle, and was taken from Fig.~\ref{Graph2}, and an estimate for the transition line from these data is shown with a dashed red line, which is a guide to the eye. Below the IMS the system will transition into the Meissner state, however we do not have direct information on $H_{\rm c1}$ from this study as the thin-plate geometry of this sample will lead to a large demagnetising effect. From values of $\kappa$ and $\lambda$ we determine later in this study, one can estimate that $H_{\rm c1}$ should be of order 100~G.

The presence of an intermediate mixed state means that BiPd must be an intermediate-kappa superconductor. Calculations find that the intermediate mixed state may appear in superconductors with $0.71 \leqslant \kappa \leqslant 1.5$~\cite{Jac71, Bra75, Bra76}.  For example, niobium has a GL parameter of 0.77 near $T_{\rm c} = 9.27$~K~\cite{Lav06}. In the next section, we will use  measurements of the VL form factor in the mixed state to obtain a value for $\kappa$ for BiPd.

\subsection{Field dependence of the form factor}

We can express the local field within the VL as a sum over the spatial Fourier components at the reciprocal lattice vectors \textbf{q}, and the magnitudes of the Fourier components $F(\mathbf{q})$ are known as the VL form factor. The magnitudes of these can be calculated from the integrated intensities $I_{\mathbf{q}}$ of Bragg reflections using the Christen formula~\cite{Chr77}:
\begin{equation}
I_{\mathbf{q}} = 2 \pi V \phi (\frac{\gamma}{4})^2 \frac{\lambda_{n}^{2}}{\Phi_{0}^{2} q} \vert F(\mathbf{q}) \vert^2,
\label{Eq1}
\end{equation}
where \textit{V} is the illuminated sample volume, $\phi$ is the neutron flux, $\gamma$ is the magnetic moment of the neutron, $\lambda_n$ is the neutron wavelength, and $\Phi_0 = h / 2e$ is the flux quantum. In our experiment, the integrated intensities $I_{\mathbf{q}}$ of the first-order Bragg reflections were calculated by fitting the rocking curves of the reflections to a Lorentzian line-shape.  These intensities were then corrected by the Lorentz factor, the cosine of the angle between the perpendicular to the rocking axis and \textbf{q}.

\begin{figure}
\begin{center}
	\includegraphics[width=0.7\linewidth]{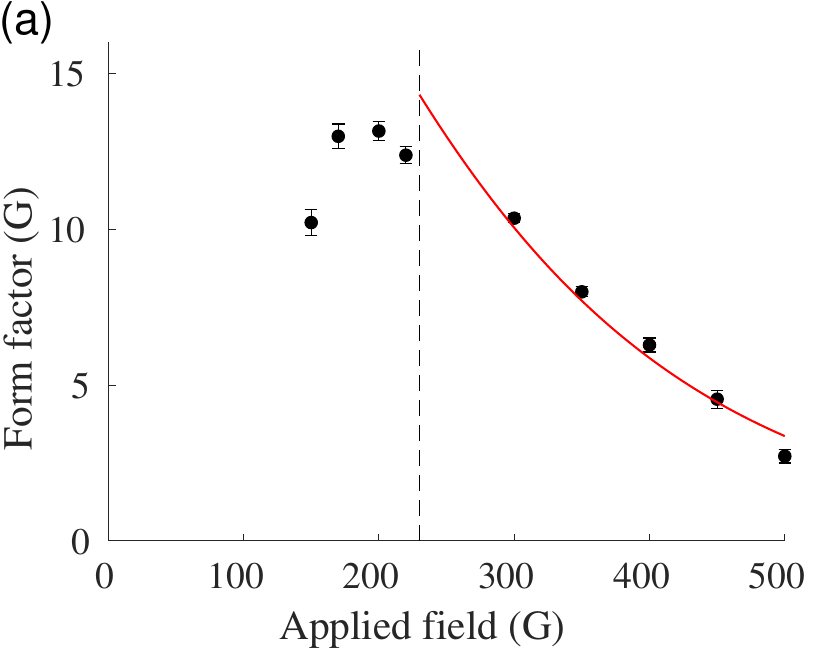}
	\includegraphics[width=0.7\linewidth]{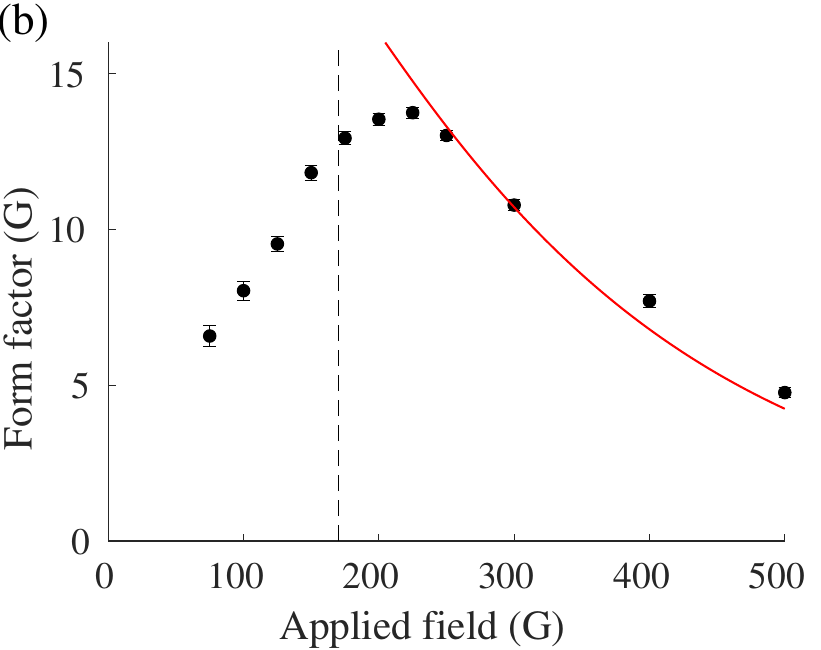}
	\includegraphics[width=0.7\linewidth]{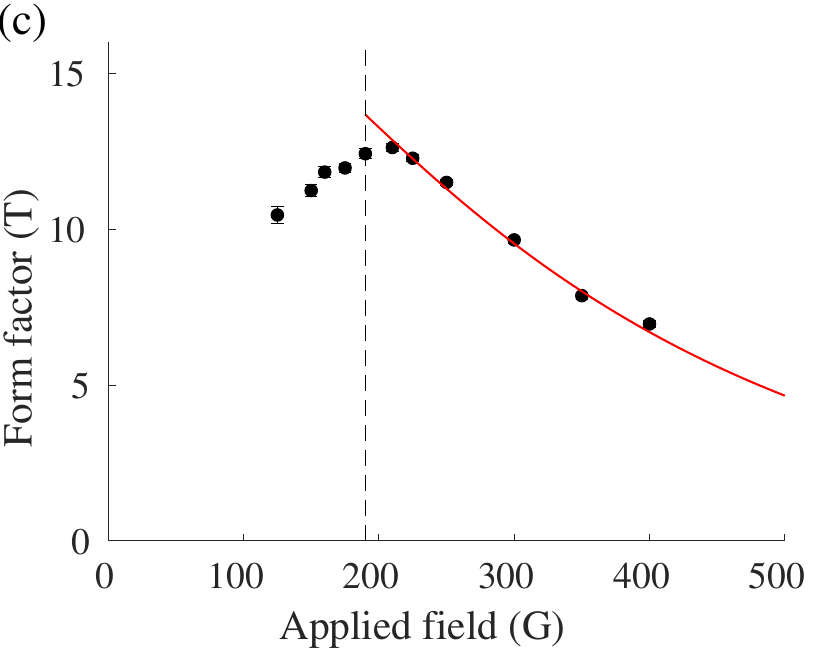}
\end{center}
	\caption{(Color online) (a) Form factor for field applied parallel to (010). (b) Form factor for field applied parallel to the $\sim$(101) direction. (c) Form factor for field applied parallel to (101). The solid red lines are fits to the extended London model, and the dashed lines indicate the transition line to the intermediate mixed state, as determined in Fig.~\ref{Graph2}.}
		\label{Graph3}
\end{figure}

Figure~\ref{Graph3} shows the form factor as a function of field for fields applied along the (010), $\sim$(101) and (101) directions, as measured at 0.1~K. The data points are an average of the form factor for all six first-order Bragg reflections of the VL. The dotted vertical line indicates the transition line between the intermediate mixed state, as determined from the scattering vector (Fig.~\ref{Graph2}), and the solid line is a fit to the extended London model, which is described below. Above the transition line, in the mixed state, the VL form factor behaves in a typical manner, decreasing with increasing field as the density of vortices increases and their spacing is reduced. However, below the transition line a completely different behaviour is observed. In this case, the intensity is rising in a reasonably linear fashion from zero. This is because the inter-vortex spacing is independent of applied field in the intermediate mixed state, but the volume-fraction of the sample occupied by vortex matter increases with field. Therefore, as the reflectivity of the vortex lattice remains constant while its volume increases, the intensity of the Bragg reflections also increases. However, this means that mixed-state methods of analysing the field-dependence of the VL form factor are no longer valid, so we discount these data from our numerical analysis.

We note that for field applied along the $\sim (101)$ and $(101)$ directions, the transition between the IMS and VL states as determined by the scattering vector and the down-turn of the form factor are separated slightly in field, by up to 50 G. The situation for field applied parallel to the (010) direction is less clear, due to a lack of data just above the IMS transition. This may be an intrinsic effect which has been seen before in niobium~\cite{Muh09}, where a hysteresis in the scattered intensity from the flux lattice was observed about the IMS transition. This has been attributed to different processes of flux intrusion and flux island formation in the presence of pinning~\cite{Ess72}. Our data was taken after field-cooling, unlike the hysteretic data from niobium, which was taken with the field changed while below $T_{\rm c}$. We might expect our field-cooled data to be equivalent to the decreasing-field data, as both are approaching the measurement point from outside the superconducting phase, but we find that it infact resembles the increasing-field data.

We fit the data within the mixed state to the `extended London model' i.e.~the London model with core corrections, given by
\begin{equation}
F(\textbf{q}) = \frac{\langle B\rangle \exp(-c q^2 \xi^2)}{1 + q^2\lambda^2},
\label{LondonFF}
\end{equation}
where $\langle B\rangle$ is the average internal induction, $q$ is the magnitude of the reciprocal space lattice vector for the first order Bragg reflection being measured, $\xi$ is the coherence length, $\lambda$ is the penetration depth, and $c$ is a parameter which accounts for the finite size of the vortex cores, with a suitable value being 0.44~\cite{Bow08, Whi11, Fur11, Cam22, Camp22}. There are some complications regarding the use of the London model to fit the data, which is that it was constructed assuming the system was far from $H_{\rm c2}$, and in the limit that the electrodynamics could be considered local in nature, restricting $\kappa \gg 1$. Since we are fitting data at low temperature, the first consideration is met, but we know that we are straining the latter as the intermediate mixed state implies that $\kappa \sim 1$. Models based on the GL theory are valid for low $\kappa$, but GL theory is only accurate close to $T_{\rm c}$, so we will retain the extended London model and consider the results to be estimates. The data were fitted using experimentally measured values of $q$, and the smooth curves displayed in the figure were calculated from a fit to these experimentally measured values of $q$ using the same relationship as in Fig.~\ref{Graph2}. We have chosen to fit to average values of the form factor, rather than trying to use the anisotropic London model to attempt to separate out values of the coherence length and penetration depth along each crystallographic axis, as out of the twelve different vortex nearest neighbour directions found within the three separate orientations of applied field, only one lies parallel to a crystal axis. This is undoubtedly due to the complications of the monoclinic structure, and prohibits the normal strategies used to fit the anisotropic London model. The fits returned values of the coherence length and penetration depth at base temperature of $\xi = 79$~nm and $\lambda = 80$~nm for $\mathbf{B} \parallel (010)$, $\xi = 73$~nm and $\lambda = 92$~nm for $\mathbf{B} \parallel \sim$(101), and $\xi = 65$~nm and $\lambda = 120$~nm for $\mathbf{B} \parallel (101)$. This gives average values of $\xi = 72$~nm and $\lambda = 98$~nm, which result in $\kappa = 1.4$. This value of $\kappa$ is consistent with the expectation, discussed in the section on the intermediate mixed state, that it should fall within the region of $0.71 \leqslant \kappa \leqslant 1.5$. However, this is much smaller than a previous result of $\kappa = 7.6$~\cite{Jos11}; we note that this value was an order of magnitude estimate derived from the higher estimate of $H_{\rm c2}$ discussed in the introduction. The presence of the IMS constrains $\kappa$, and so this indicates that the higher estimate of $H_{\rm c2}$ cannot be correct.

\subsection{Temperature dependence of the form factor}

The temperature dependence of the superfluid density, $\rho_s (T)$, which is accessible through the form factor, is one method to test the gap structure and thus gain information about the pairing state. In this case, the amount of thermal energy required to break Cooper pairs is determined by the magnitude of the gap, and so the form of the curve $\rho_s (T)$ is fundamentally determined by the gap. Typically, superconductors which have a full gap arising from \textit{s}-wave pairing have a constant low-temperature value of the superfluid density as the thermal energy falls below the gap value on approach to absolute zero. Conversely, superconductors which have nodes in the gap can experience quasiparticle excitations even at the lowest temperatures, and so have a finite slope of $\rho_s (T)$ at zero temperature. One of the key predictions of the possible mixing of \textit{s}- and \textit{p}-wave pairing states in NCS superconductors is the emergence of either anisotropic gaps or accidental nodes, which would modify the form of $\rho_s (T)$ away from that expected for centrosymmetric superconductors. A theory to model this has been constructed by Hayashi~\textit{et al.}~\cite{Hay06}.  This has been used successfully to identify mixed gaps in Li$_2$Pd$_3$B and Li$_2$Pt$_3$B~\cite{Yua06}, and Ru$_7$B$_3$~\cite{Cam22}. The data presented here demonstrate neither the flat region at low temperatures characteristic of \textit{s}-wave superconductors nor the pronounced finite slope at zero temperature of nodal superconductors~\cite{Pro06}, but rather a slight slope down to the lowest temperatures in a similar manner to the data from Ru$_7$B$_3$~\cite{Cam22}. We therefore employ the same model here in order to determine if the data supports NCS gap mixing in BiPd, and if so, to what extent.

In the mixed state, the magnitude of the scattering vector, which contributes to the form factor, is determined solely by the applied field and the vortex lattice structure. Therefore, if measuring at a fixed field, and in the absence of any VL structure transitions, the change in the form factor comes only from changes in the superfluid density. These are the conditions to be used to obtain $\rho_s (T)$. 

\begin{figure}
	\includegraphics[width=0.7\linewidth]{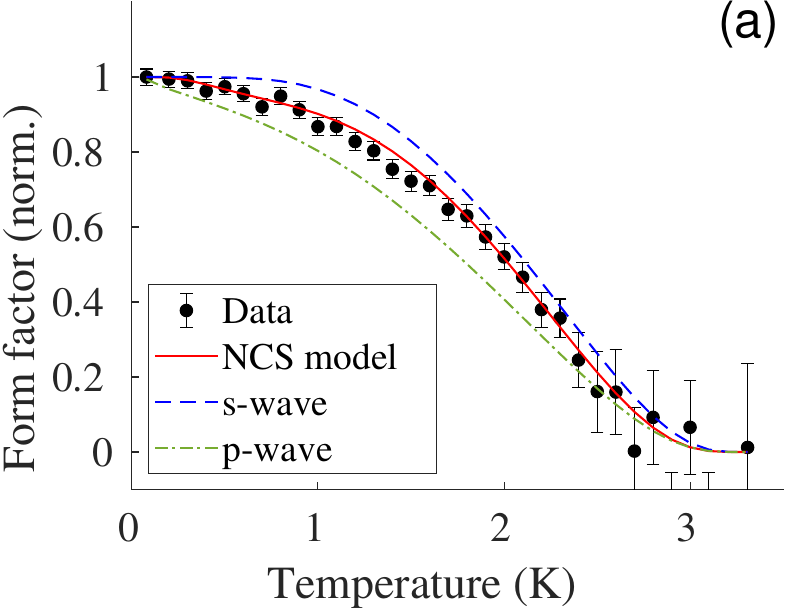}
	\includegraphics[width=0.7\linewidth]{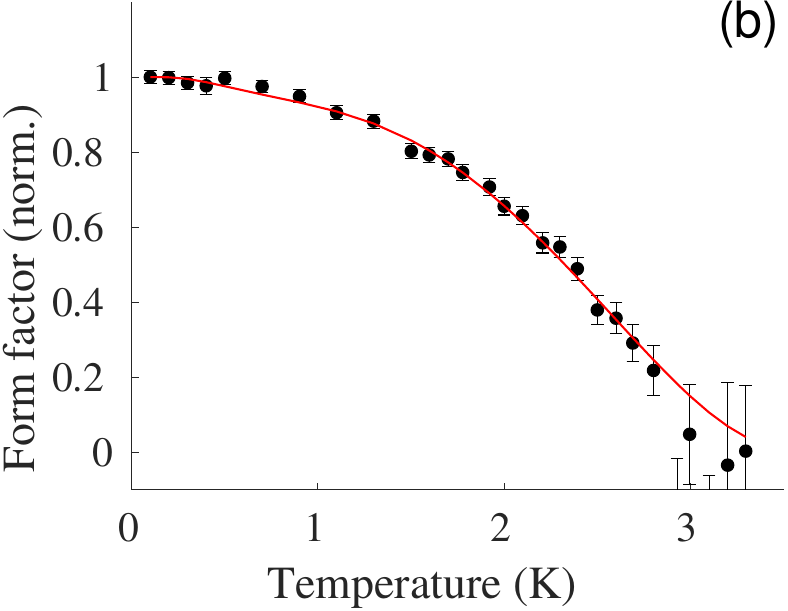}
	\includegraphics[width=0.7\linewidth]{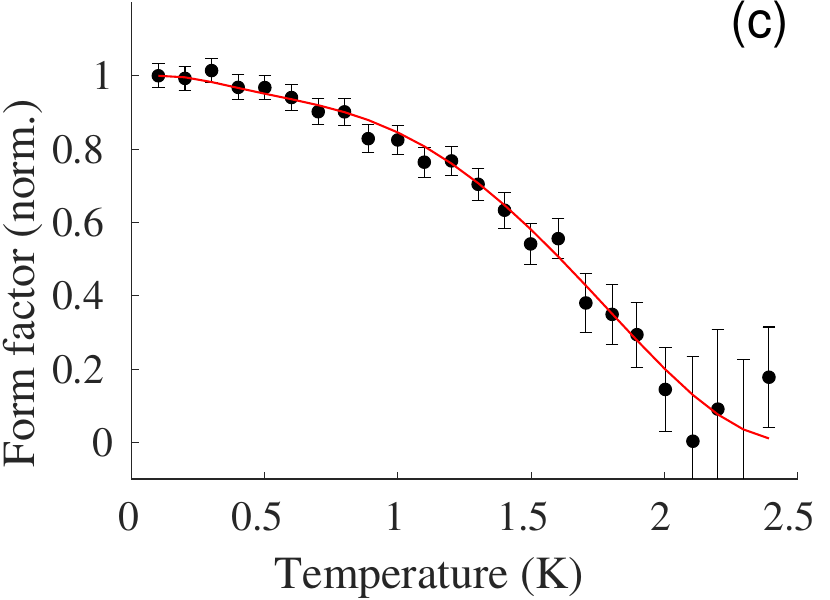}
	\includegraphics[width=0.7\linewidth]{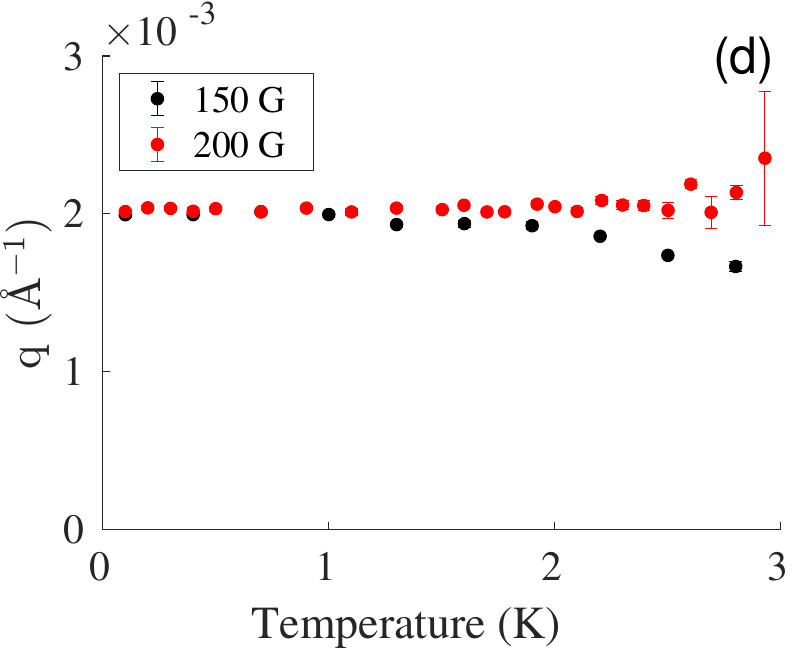}
	\caption{(Colour online) Temperature dependence of the form factor, normalised to its low-temperature value, for several different orientations and temperatures. (a) Field parallel to the (010) direction at 250~G. (b) Field parallel to the $\sim$(101) direction at 200~G. (c) Field parallel to the $\sim$(101) direction at 400~G. Solid lines are a fit to the NCS model described in the text, and panel (a) also has plots of \textit{s}-wave and \textit{p}-wave models. (d) Temperature dependence of the wavevector $q$ of one of the first order Bragg reflections from the vortex lattice, as measured at a field above $B_{\rm IMS}$ at 200 G (red), and below $B_{\rm IMS}$ at 150 G (black), with the field applied along the $\sim$(101) direction.}
		\label{Graph4}
\end{figure}

However, in the IMS the scattering vector is not determined by the applied field, but rather the balance between the attractive and repulsive parts of the inter-vortex interaction. This balance can change with temperature, and modifies both \textbf{q} and the scattered intensity from the vortex lattice, both of which contribute to the form factor. This has been seen quite clearly in, for instance, a SANS study on the vortex lattice in niobium~\cite{Bac19}. We do see a change in the scattering vector with temperature in the IMS of BiPd, which is shown in Fig.~\ref{Graph4}(d). While the scattering vector of the data from the mixed state remains constant as a function of temperature, the scattering vector of the data taken within the intermediate mixed state decreases as the temperature increases. This is due to decreasing $B_{\rm 0}$ within the vortex islands, as the vortices increase in size with increasing temperature. In order to avoid the complications of multiple contributing parameters varying within a single measurement we confine ourselves to temperature scans of the VL in the mixed state.

Equation~\ref{LondonFF} may be used to give the form factor as a function of temperature at a fixed field. Two parameters in this relationship have a temperature dependence: the penetration depth $\lambda$ and the coherence length $\xi$, so we can use this to fit the temperature dependence of the form factor. The NCS model of Hayashi~\textit{et al.} gives the superfluid density as a function of temperature, which enters the form factor through the penetration depth as per the relationship $1 / \lambda^{2} \: \propto	\: \rho_{s}$. The gap also contributes to the form factor through the coherence length $\xi$, which is inversely proportional to the superconducting gap given by the model. This involves Cooper pairing including both spin singlet and spin triplet contributions to the order parameter. From this, an expression for the superfluid density was calculated, which describes the system in terms of the ratio of the singlet ($\Psi$) and triplet ($\Delta$) order parameters, $\nu = \Psi/ \Delta$. It also involves the difference of the density of states of the spin-split Fermi surfaces, labelled I and II, $\delta = (N_{\rm I} - N_{\rm II})/(N_{\rm I} + N_{\rm II})$, and finally the coupling constant $\lambda_{\rm m}$ which describes Cooper pair scattering between the two channels. We calculate the temperature dependence of the superfluid density and the gap by the method described by Hayashi \emph{et al.}~\cite{Hay06}. 

For comparison, we also fitted the data to both fully gapped and nodal models appropriate for a superconductor with no order parameter mixing, in the same manner as was done in previous studies of the VL form factor by SANS~\cite{Fur11, Whi11, Mor14, Cam22, Camp22}. We have chosen to use \textit{s}-wave and \textit{p}-wave models as these are the two potential components of the order parameter in NCS superconductors. For these cases, the superfluid density in the clean limit is given by the expression
\begin{equation}
\begin{split}
\rho_{s}(T) & = 1 - \frac{1}{4\pi k_{B}T} \\
& \int_{FS} \int_0^\infty \cosh^{-2} \left( \frac{\sqrt{\varepsilon^{2} + \Delta_{\textbf{k}}^{2}(T,\theta,\phi)}}{2k_{B}T} \right) \mathrm{d}\theta \mathrm{d}\phi \mathrm{d}\varepsilon .
\end{split}
\end{equation}
Here $\Delta_{\textbf{k}}$ describes the gap function, $\theta$ and $\phi$ are the polar and azimuthal angles about the Fermi surface (FS) and $\sqrt{\varepsilon^{2} + \Delta_{\textbf{k}}^{2}(T,\theta,\phi)}$ gives the excitation energy spectrum. The gap is separated such that its momentum and temperature parts can be described by $\Delta_{\textbf{k}}(T,\theta,\phi) = g_{\textbf{k}}(\theta, \phi)\Delta_{0}(T)$, where $g_{\textbf{k}}(\theta, \phi)$ is the momentum dependence of the gap as a function of angle around the Fermi surface. The temperature dependence of the gap can be approximated by
\begin{equation}
\Delta_{0}(T) = \Delta_{0}(0) \tanh \left( \frac{\pi T_{\rm c}}{\Delta_{0}(0)}\sqrt{a \bigg ( \frac{T_{c}}{T} - 1 \bigg ) } \right),
\end{equation}
where ${\Delta_{0}}$ is the magnitude of the gap at zero temperature and $a$ is a parameter related to the pairing state~\cite{Pro06}. For \textit{s}-wave pairing, we took  $g_{\textbf{k}}(\theta, \phi) = 1$, $\Delta_{0}(0) = 1.76 T_{\rm c}$ and $a = 1$, while for \textit{p}-wave pairing $g_{\textbf{k}}(\theta, \phi) =  \cos (\phi)$, $\Delta_{0}(0) = 2.14 T_{\rm c}$ and $a = 4/3$. These latter values of $\Delta_{0}(0)$ and $a$ are those used for \textit{d}-wave pairing~\cite{Pro06}, and for  $g_{\textbf{k}}(\theta, \phi)$ we have used a two-dimensional approximation. These were chosen to capture the essential features of the nodal pairing state. In the NCS case, the fitting parameters were given by the theory~\cite{Hay06}.

In Fig.~\ref{Graph4}, we present the temperature dependence of the form factor, alongside fits to the noncentrosymmetric model of Hayashi~\textit{et al.}~\cite{Hay06} for several different magnitudes and orientations of the applied magnetic field. In panel (a), the field is parallel to the (010) direction at 250 G, and in panels (b) and (c) are fields of 200 G and 400 G respectively parallel to the $\sim$(101) direction. In panel (a) we show examples of the the pure $s$- and $p$-wave models; neither model describes the data well. We found no set of parameters for which either of these models could describe the data, indicating that neither a fully gapped nor nodal description is suitable. For the fit to the non-centrosymmetric model, we held the parameter $\lambda_{\rm m}$ constant at 0.2, as suggested by Hayashi \emph{et al.}~\cite{Hay06}.  This value has been used in other examples of the implementation of this model ~\cite{Yua06, Cam22}, as it was determined in the initial publication that varying this parameter does not have any strong quantitative or qualitative effects on the model. Following this, we found a solution with common values for $\nu$ and $\delta$ for all of the data sets. The fits returned values of $\nu = 1.45 \pm 0.1$ and $\delta = 0.75 \pm 0.05$.

These values clearly indicate that the mixed order parameter of the NCS model is an appropriate description for these data, while neither a single-gap \textit{s}-wave or nodal order parameter would fit. A value of $\delta \neq 1$ indicates a difference in the density of states between the spin-split bands of an NCS superconductor, and the value of $\nu = 1.45$ indicates that a reasonable amount of order-parameter mixing is taking place, leading to a nodeless but anisotropic gap. This result is in reasonable consistency with the published data, and offers a possible resolution to the conflict between different results discussed in the introduction. We have a multi-gap scenario due to spin-split bands from the ASOC, with anisotropic but nodeless gaps from singlet-triplet mixing with the singlet component dominant. The gap on each of the spin split bands is also different, with one being described by $\mid \psi + \Delta {\rm sin} \theta \mid$, and the other by $\mid \psi - \Delta {\rm sin} \theta \mid$~\cite{Hay06}. This is consistent with the multiple reports of a two-gap order parameter~\cite{Min12, Ram17, Jia14}, but may also account for results which reported conventional behaviour~\cite{Pee14, Sun15}, as we still have a finite gap over all the Fermi surface.

\begin{figure}
	\includegraphics[width=1\linewidth]{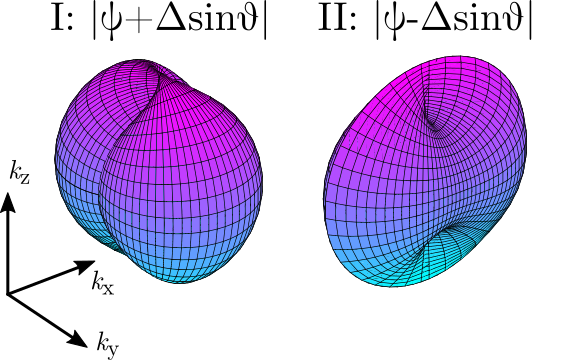}
	\caption{(Color online) Plot of the gap values on two spin-split Fermi surfaces from the NCS fit described in the text.}
		\label{Graph5}
\end{figure}

Using the value of $\nu = 1.45$ from the fits in Fig.~\ref{Graph4}, we plot the modulus of the gap values on each of the spin-split Fermi surfaces, I and II, in Fig.~\ref{Graph5}. This was done using the method described in Ref.~\cite{Hay06}. The gap on Fermi surface I is given by $\mid \psi + \Delta {\rm sin} \theta \mid$, and the gap on Fermi surface II is given by $\mid \psi - \Delta {\rm sin} \theta \mid$, where $\theta$ is the azimuthal angle in momentum space. The gaps are reasonably anisotropic, as expected for a value of $\nu$ close to 1, but at no point do they fall to zero as the singlet component is larger in magnitude than the triplet component.

In Table~1 we present values of $\nu$ and $\delta$ from fitting the temperature dependence of the superfluid density to the NCS model, alongside values from other NCS superconductors where this model has also been used. The values of the energy of the ASOC are also presented, following Smidman \textit{et al.}~\cite{Smi17}. We see some general trends in the data with regards to the ratio of singlet to triplet gap, $\nu$, where systems with larger ASOC tend to have more prevalent triplet components. It was noted in the paper on Li$_2$Pd$_3$B and Li$_2$Pt$_3$B that the ratio of the density of states, $\delta$, was expected to be proportional to the ASOC~\cite{Yua06}, and for the two materials they investigate this is true, although it does not hold for the others as they have larger values of $\delta$ with a significantly lower ASOC than the platinum compound. However, we note that unlike the platinum and palladium compounds the other materials investigated are not related to each another.

\begin{center}
\begin{table}
\begin{tabular}{ |c|c|c|c| } 
 \hline
 Material & $\nu$ & $\delta$ & E$_{\rm ASOC}$ (meV) \\ 
 \hline\hline
 Li$_2$Pd$_3$B~\cite{Yua06} & 4 & 0.1 & 30 \\ 
 Li$_2$Pt$_3$B~\cite{Yua06} & 0.6 & 0.3 & 200\\
 Ru$_7$B$_3$~\cite{Cam22} & 2.5(5) & 0.70(7) & 20\\ 
 BiPd & 1.45(1) & 0.75(5) & 50 \\
 \hline
\end{tabular}
\label{Table1}
\caption{Values for the singlet-to-triplet gap ratio, $\nu = \Psi/\Delta$, and the density of states ratio, $\delta$, as measured from NCS materials whose superfluid density as a function of temperature has been fitted to the model described in the text. Values for the energy of the ASOC are also given, taken from Ref.~\cite{Smi17}.}
\end{table}
\end{center}

\section*{Conclusions}

We have observed the presence of the intermediate mixed state in BiPd, which means that the material must be a low-$\kappa$ Type-II superconductor, with a low value of the upper critical field in the bulk of the sample. The values of the coherence length and penetration depth that we estimate from a London-model fit to the field dependence of the VL form factor conform to this expectation, giving a value of $\kappa \sim 1.4$. Further, we fitted the temperature dependence of the VL form factor to a model derived from the mixed singlet-triplet order parameter of non-centrosymmetric superconductors, and find the data to be in good agreement with the model. This work may also motivate a thorough theoretical study of the richness of the vortex-matter phase diagram and the structure of the VL in non-centrosymmetric superconductors with spin-triplet mixing.

\section*{Acknowledgements}
ASC, ATH, EMF and EB acknowledge support from the UK Engineering and Physical Sciences Research Council grant numbers EP/J016977/1 and EP/G027161/1. LL acknowledges financial support from the Institut Laue-Langevin and the University of Birmingham. ASC acknowledges support from Lund University during the preparation of the manuscript. This work is based on experiments performed at the Swiss spallation neutron source SINQ, Paul Scherrer Institute, Villigen, Switzerland. Experiments at the ILL gave data available at DOI 10.5291/ILL-DATA.5-42-374~\cite{ILL1} and 10.5291/ILL-DATA.5-42-390~\cite{ILL2}.

%

\end{document}